\documentclass[11pt]{article}
\usepackage{psfig,float}

\newcommand{\rbibitem}[1]{\bibitem{#1}}
\newcommand{\be}{\begin{equation}}
\newcommand{\ee}{\end{equation}}
\newcommand{\ba}{\begin{eqnarray}}
\newcommand{\ea}{\end{eqnarray}}

\renewcommand{\mathrm}[1]{{\rm #1}}

\begin{document}
\begin{titlepage}
\begin{flushright}
{FTUV/98-36}\\
{IFIC/98-37}\\
\end{flushright}
\vspace{2cm}
\begin{center}
{\large\bf 
$K {\bar K}$ scattering amplitude to one loop in chiral perturbation 
theory, its unitarization and pion form factors} \\
\vfill
{\bf Francisco Guerrero and Jos\'e Antonio Oller}\\[0.5cm]
 Departament de
 F\'{\i}sica Te\`orica, Universitat de Val\`encia and\\
 IFIC, CSIC - Universitat de Val\`encia,
 C/ del  Dr. Moliner 50, \\ E-46100 Burjassot (Val\`encia),
Spain\\[0.5cm]
\end{center}
\vfill
\begin{abstract}
We have calculated the $K {\bar K} \rightarrow K {\bar K}$ 
scattering amplitude to next to leading order in Chiral Perturbation
Theory. Then, making use of a unitarization procedure with one or several
coupled channels ($\pi\pi$,$K {\bar K}$ in our case) we have calculated
the $\pi\pi \rightarrow \pi\pi$, $\pi\pi \rightarrow K {\bar K}$ and
$K {\bar K} \rightarrow K {\bar K}$ S and P waves in good 
agreement with the experiment up to $\sqrt{s} \simeq 1.2$ GeV. 

The $\pi\pi$ scattering lengths with isospin and spin (I,J) equal to
(0,0), (1,1) and (2,0) are also 
calculated in agreement with experiment and former Chiral Perturbation 
Theory calculations.

Finally we have employed
these amplitudes, making use of an Omn\`es representation, to 
calculate the scalar and the vector pion form factors, obtaining a 
good agreement with the available experimental data.
\end{abstract}
\vspace*{1cm}
\vfill
May 1998
\end{titlepage}

\vspace{0.5cm}
\noindent{\bf 1. \, Introduction}
\vspace{0.5cm}

The low energy effective theory of the strong interactions (QCD) is
Chiral Perturbation Theory (ChPT). The chiral symmetry constraints
\cite{W1} are a powerful tool, enough to determine the low energy matrix
elements in a systematic way \cite{G1,G2,P1,E1,M1}. In this way one can 
evaluate the scattering amplitudes corresponding to the lightest octet of 
pseudoscalar mesons ($\pi, K, \eta$) in a loop expansion. This has already 
been done for the $\pi\pi \rightarrow \pi\pi$ scattering amplitude for the 
$SU(2)$ case in \cite{G1} and extended to $SU(3)$ in \cite{M2}. 
The $K \pi \rightarrow K \pi$ scattering amplitude is also evaluated 
in \cite{M2}. 

This loop expansion can also be seen as an expansion in powers of the masses 
and external tetramomenta of the pseudoscalars
over some typical scale around 1 GeV, so that 
the series is, in principle, only useful for 
low energy. Furthermore, in the meson-meson scattering, it is common 
the appearance of resonances with a typical mass around 1 GeV, which obviously 
can not be reproduced in a power expansion because they correspond to poles in 
the T matrix. The region of validity of the series depends on the 
channel (I,J) considered, where I accounts for the isospin and J for the 
angular momentum. A way to get a hint about this convergence problem is to 
calculate the next to leading contribution (${\cal O} (p^4)$) and compare 
it with the lowest order contribution(${\cal O} (p^2)$). 
This seems logical and is the 
reason advocated in \cite{M2} to justify their $K \pi$ ${\cal O}(p^4)$ 
calculation 
even though the threshold for this reaction is 0.64 GeV and typically the 
chiral expansions break around $\sqrt{s} \sim 0.5-0.6$ GeV.

In this paper we first calculate the $K {\bar K} \rightarrow K {\bar K}$ 
scattering amplitude at ${\cal O} (p^4)$ in ChPT. 
This implies to evaluate two 
amplitudes due to the fact that both I=0,1 are possible. In this way we 
calculate the $K^+ K^- \rightarrow K^+K^-$ and $K^+ K^- \rightarrow K^0
{\bar K^0}$ scattering amplitudes. It is not necessary to calculate the
$K^0 {\bar K^0} \rightarrow K^0 {\bar K^0}$ amplitude because in the isospin 
limit $T(K^+ K^- \rightarrow K^+ K^-)=
T(K^0 {\bar K^0} \rightarrow K^0 {\bar K^0})$.

This calculation seems in 
principle to be unreasonable since the $K {\bar K}$ threshold is almost 
1 GeV, and furthermore, for I=0,1 at this energy the $f_0 (980), a_0 
(980)$ 
resonances appear which couple strongly to the $K {\bar K}$ channel. In fact,
after making this ${\cal O} (p^4)$ calculation, 
we will see that, at this energy, 
the ${\cal O} (p^4)$ result is larger than the ${\cal O} (p^2)$ 
lowest order 
calculation. Obviously, what all this is telling us is that one cannot 
rely in the perturbative chiral expansion at these energies. 
This implies that one cannot compare the predictions directly with the 
experiment, using 
the ${\cal O} (p^2)$ plus ${\cal O} (p^4)$ ChPT amplitudes.

In the last years, there have been several attempts to extend ChPT to higher 
energies. One of them uses an effective field theory with explicit resonance 
fields as degrees of freedom \cite{E2,E3}. This resonance chiral effective theory 
has one drawback: applying only the constraints coming from the symmetry the 
number of free parameters grows so fast with higher orders that in practice 
predictions become impossible.
However, using this lagrangian at ${\cal O}(p^4)$ and imposing some short 
distance QCD constraints \cite{E3} one can obtain interesting and good 
results. For example, the study of the vector pion form factor done in  
\cite{Gu1}.


Another attempt (the one we are going to use here) is to unitarize the ChPT 
amplitudes. In \cite{T1,D1,D2} the Inverse Amplitude Method (IAM) 
was developed and 
used to study the $\pi\pi$ and $K \pi$ scattering. This method has 
the problem that only 
elastic unitarization is employed, so that, for those cases where inelastic 
thresholds are important, as the $K {\bar K}$ one in $I=J=0$, the improvement 
over ChPT is poor. But for those other channels predominantly elastic, as the 
$I=J=1$, the improvement is remarkable and, in fact, in this channel 
the $\rho$ resonance is obtained. In order to use the IAM 
one needs the ${\cal O} (p^4)$ ChPT amplitude. More recently, in \cite{O1}, 
a resummation 
of the chiral series was done inspired in the Bethe-Salpeter (BS) 
equations 
for the $J=0$ sector making use only of the ${\cal O} (p^2)$ ChPT 
amplitudes
and imposing unitarity with coupled channels, $\pi\pi$ and $K {\bar K}$. The 
agreement with the $I=J=0$ and $I=1$, $J=0$ experimental data was very good, 
reproducing 
the presence of the $f_0 (980)$ and $a_0 (980)$ resonances respectively. 
However, with this 
method the higher orders are obtained through unitarity loops in the s 
channel  which are subleading in the large $N_c$ limit and then fails in the 
vector sector where the large $N_c$ limit works very well. The leading part in 
this latter limit are tree level contributions  appearing through the chiral 
lagrangians. In fact in \cite{M2} it is seen that in the $J=1$ $K \pi$
scattering the 
${\cal O} (p^4)$ corrections are clearly dominated by the polynomial 
counterterms contribution coming from the ${\cal O} (p^4)$ chiral lagrangian.
This same phenomenon is also seen in the P-wave $\pi\pi$ scattering and in the 
vector form factor \cite{G1,Gu1,G4,Do1,F2}.
It is clear then that a unitarization procedure which could include the 
success of the IAM and BS approaches, or equivalently, a method that could 
handle unitarization in coupled channels incorporating both the 
${\cal O} (p^2)$ 
and ${\cal O} (p^4)$ chiral amplitudes would be very welcome.
This in fact has been recently done in \cite{O2} 
and the resonances appearing in both $J=1$ ($\rho (770)$ and $K^* (890)$) and 
$J=0$ ($a_0 (980)$, $f_0 (980)$) channels were reproduced. This is in fact the 
method we are going to use here. The novelties are that while in \cite{O2} the 
${\cal O} (p^4)$ chiral amplitudes were approximated in a way inspired in 
\cite{O1}, here we are going to use the exact $\pi\pi$ and $K \pi$ scattering 
amplitudes \cite{G1,M2} 
and the $K {\bar K}$ ones, which we calculate here to ${\cal O} (p^4)$. 
With these ingredients  
we obtain the scattering amplitudes for the channels (I,J)=(0,0), (1,1) and 
(2,0), generating dynamically the $f_0$ and $\rho$ resonances.
The agreement with experiment is good up to around $\sqrt{s} \simeq $ 1.2 GeV
as it will show. 


Finally, making use of the calculated $\pi\pi \rightarrow \pi\pi$ phase shifts 
we derive an Omn\`es representation \cite{Mu1,Om1} for the scalar and vector 
pion form factors based on the Watson final state theorem 
\cite{Wa1}. For the case of the vector form factor 
experimental data are available and, with our calculation, 
the agreement is rather good. In the case of the scalar 
form factor we compare the calculation making the unitarization with and 
without including the $K {\bar K}$ channel and the differences are very 
significative, even at energies around $\left | \sqrt{s} \right |=500$ MeV. 
The $f_0 (980)$ resonance appears clearly in the scalar form factor of the 
pion.

\vspace{0.5cm}
\noindent{\bf 2. \, The $K {\bar K} \rightarrow K {\bar K}$ scattering 
amplitude to next to leading order in ChPT}
\vspace{0.5cm}

As it was stated before, in order to calculate the $K {\bar K}$ amplitude, we 
need to evaluate two amplitudes. These two independent isospin 
amplitudes cannot be connected by crossing symmetry
because they have different absolute values for the strangeness. 
This is contrary to what
happens in $K \pi$ scattering with $I=3/2$, $I=1/2$.

We calculate the amplitudes $K^+ K^- \rightarrow K^+ K^-$ and $K^+ K^- 
\rightarrow K^0 {\bar K^0}$ which we denote by $T_{cc}$ and 
$T_{cn}$ respectively.

The scattering amplitudes with definite isospin 
$T^{(I)}$ can be written in terms of $T_{cc}$ and $T_{cn}$ in the following 
way:

\begin{eqnarray}
&T^{(0)}(s,t,u)=& T_{cc} (s,t,u) + T_{cn} (s,t,u)  \nonumber  \\
&T^{(1)}(s,t,u)=& T_{cc} (s,t,u) - T_{cn} (s,t,u)
\end{eqnarray}
  
\noindent
We now proceed to describe the calculation scheme for these amplitudes up to 
${\cal O} (p^4)$.

At lowest order one has the ChPT lagrangian at ${\cal O} (p^2)$ 

\be
{\cal L}_2=\frac{f_0^2}{4} \left < \partial_{\mu} U^{\dagger} 
\partial^{\mu} U +
{\cal M} \left(U+U^{\dagger}\right) \right >
\ee

\noindent
where $\left< \right>$ stands for the trace of the $3 \times 3$ 
matrices built from $U (\Phi)$ and 
$\cal M$,

\be
U(\Phi)=\exp \left( \frac{i \sqrt{2}}{f_0} \Phi \right)
\ee

\noindent
where $\Phi$ is expressed in terms of the Goldstone boson fields as

\be
\Phi (x)= \left[ \matrix{\frac{1}{\sqrt{2}}\pi^0+\frac{1}{\sqrt{6}}\eta & 
\pi^+ & K^+ \cr \pi^- & -\frac{1}{\sqrt{2}}\pi^0+\frac{1}{\sqrt{6}}\eta & K^0 
\cr K^- & {\bar K^0} & -\frac{2}{\sqrt{6}}\eta} \right]
\label{phi}
\ee

\noindent
The mass matrix ${\cal M}$ is given by 

\be
{\cal M}=\left[ \matrix{{\hat m}^2_{\pi} & 0 & 0 \cr 0 & {\hat m}^2_{\pi} & 0 
\cr 0 & 0 & 2 {\hat m}^2_K - {\hat m}^2_{\pi} } \right]
\ee

\noindent
in the isospin limit. Where ${\hat m}$ means bare masses.

From this lagrangian we can evaluate the lowest order contribution to the 
$K {\bar K}$ scattering amplitudes

\begin{eqnarray}
&T_{cc,2}(s,t,u)&=\frac{1}{f^2_0} \left[ \frac{2}{3} 
{\hat m}^2_K + \frac{4}{3} 
m^2_K - u \right] \nonumber \\
&T_{cn,2}(s,t,u)&=\frac{1}{2f^2_0} \left[ \frac{2}{3} {\hat m}^2_K + 
\frac{4}{3} m^2_K -u \right]
\end{eqnarray}

\noindent
where the subindex 2 means ${\cal O} (p^2)$.

At ${\cal O} (p^2)$ $f_0=f_{\pi}=93.3$ MeV and ${\hat m}_K=m_K=495.7$ MeV. 
But when we 
go to next order these equalities do not hold. This is the reason why we keep 
the distinction between bare and physical masses.

\begin{figure}[ht]
\centerline{\protect\hbox{
\psfig{file=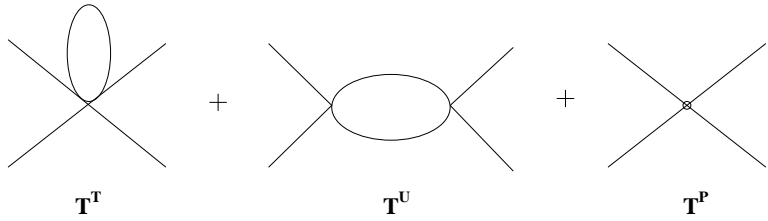,width=0.8\textwidth,silent=}}}
\caption{Diagrams at ${\cal O} (p^4)$}     \label{diagram}
\end{figure} 

At ${\cal O} (p^4)$ one has to calculate the diagrams schematically shown 
in Fig. 1. $T_4^T$ represents contributions coming from the ${\cal L}_2$ ChPT 
lagrangian with six 
fields and a tadpole loop. The $T_4^U$ represents the loops constructed from 
the ${\cal L}_2$ amplitudes with four fields appearing in the vertices of the 
loop. We will call this contribution unitarity loops because it makes the 
amplitude unitary at ${\cal O} (p^4)$. These loops include contributions 
from loops in the s,t and u channels, as shown in Fig. 2.

\begin{figure}[ht]
\centerline{\protect\hbox{
\psfig{file=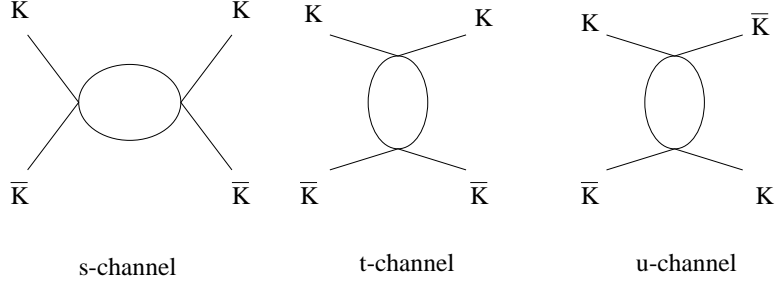,width=0.8\textwidth,silent=}}}
\caption{Diagrams for the s, t and u channels}     \label{channel}
\end{figure}

Finally the $T_4^P$ amounts for the ${\cal O} (p^4)$ polynomial contribution 
coming for the ${\cal L}_4$ ChPT lagrangian, which can be written as,

\begin{equation}
{\cal L}_4=L_1 \left<\partial_{\mu} U^{\dagger} \partial^{\mu} U 
\right>^2+
L_2 \left<\partial_{\mu} U^{\dagger} \partial_{\nu} U 
\right>\left<\partial^{\mu} U^{\dagger} \partial^{\nu} U \right>
\end{equation}
\begin{center}
$+L_3 \left<\partial_{\mu} U^{\dagger} \partial^{\mu} U \partial_{\nu} 
U^{\dagger} \partial^{\nu} U \right>+L_4 \left<\partial_{\mu} 
U^{\dagger} 
\partial^{\mu} U \right> \left<U^{\dagger} {\cal M}+{\cal M}^
{\dagger} U \right>$
\end{center}
\begin{center}
$+L_5 \left<\partial_{\mu} U^{\dagger} \partial^{\mu} U 
\left(U^{\dagger} {\cal 
M}+{\cal M}^{\dagger} U \right)\right>+ L_6 \left<U^{\dagger} {\cal M}
+{\cal M}^{\dagger} U \right>^2$
\end{center}
\begin{center}
$+L_7 \left<U^{\dagger} {\cal M}-{\cal M}^{\dagger} U \right>^2+L_8 
\left<{\cal M}^{\dagger} U {\cal M}^{\dagger}U +U^{\dagger} {\cal M} 
U^{\dagger} {\cal M} \right>$
\end{center}

\noindent
When taking into account the wave function renormalization, Fig. 3, and the 
relation between bare and physical masses and decay constants, other 
${\cal O}(p^4)$ contributions come from the lowest order amplitudes.

\begin{figure}[ht]
\centerline{\protect\hbox{
\psfig{file=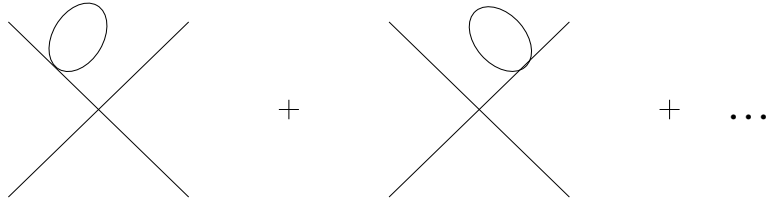,width=0.8\textwidth,silent=}}}
\caption{Wave function renormalization}     \label{ren}
\end{figure}

The relation between $m_K$ and ${\hat m}_K$ and the one between $f_0$ and 
$f_{\pi}$ at ${\cal O} (p^4)$ can be obtained from \cite{G2}.

The final amplitude up to next to leading order is then obtained summing all 
the former 
contributions. We express it divided in four parts: $T_2$, $T_4^T$, $T_4^U$ 
and $T_4^P$ where the subindices indicate the order in powers of momentum.

In $T_4^T$ and $T_4^P$ we have also included, in addition to the one coming 
from Fig. 1, the 
${\cal O} (p^4)$ contributions from the renormalization of wave function, 
masses and decay constants. The ones with $L_i$ parameters
are included in $T^P_4$ and the rest, of tadpole type, in $T^T_4$.

The amplitude for $K^+ K^- \rightarrow K^+ K^-$ is:

\be
T_{cc,2} (s,t,u)=\frac{2 m^2_K -u}{f_{\pi}^2}
\ee

\begin{eqnarray}
&T^T_{cc,4}=&\frac{A^r_{\pi}}{288 f_{\pi}^4 {\pi}^2} 
(8m^2_K+m^2_{\pi}-3u)+
\frac{A^r_K}{288 f_{\pi}^4 {\pi}^2} (8m^2_K+3u)  \nonumber \\
&           &+\frac{A^r_{\eta}}{288 f_{\pi}^4 {\pi}^2} 
(20 m^2_K-m^2_{\pi}-6u)
\end{eqnarray}

\begin{eqnarray}
&T^U_{cc,4}&=
\frac{-20m^4_K-2s^2-2su+u(3m^2_{\pi}+u)+m^2_K (8s+11u-4m^2_{\pi}
)}{192 f^4_{\pi} {\pi}^2}   \nonumber \\
&          &+\frac{A^r_{\pi}}{288 f^4_{\pi} {\pi}^2} 
(13 m^2_K-m^2_{\pi}-6u)+ 
\frac{A^r_K}{576 f^4_{\pi} {\pi}^2} (32 m^2_K -15u) \nonumber \\
&          &+\frac{A^r_{\eta}}{576 f^4_{\pi} {\pi}^2} (22 m^2_K + 
2 m^2_{\pi}-15 u)   \nonumber  \\
&          &+\left\{ \frac{B^r_{\pi} (s)}{1536 f^4_{\pi} {\pi}^2} 
\left[ 8m^2_K 
(s-4m^2_{\pi})+s(7s-4u)+8m^2_{\pi} (s+2u) \right] \right. 
\nonumber    \\
&          &
+\frac{B^r_K (s)}{384 f^4_{\pi} {\pi}^2} 5 (8m^4_K+s(s-u)-4m^2_K t) 
\nonumber \\ &&+\frac{B^r_{\eta} (s)}{4608 f^4_{\pi} {\pi}^2} (8m^2_K-9s)^2   
\nonumber   \\
&          &+ \left. \frac{B^r_{\pi \eta} (s)}{768 f^4_{\pi} {\pi}^2} 
(4m^2_K-3s)^2 + \, \, s \leftrightarrow t \right\}
+ \frac{B^r_K (u)}{32 f^4_{\pi} {\pi}^2} (u-2 
m^2_K)^2
\end{eqnarray}

\begin{eqnarray}
&T^P_{cc,4}&=L_1 \frac{8}{f^4_{\pi}} (-8m^4_K+s^2+t^2+4 m^2_K u) 
\nonumber \\ && +L_2 \frac{4}{f^4_{\pi}} (s^2+t^2+2u^2-4 m^2_K u) 
\nonumber \\&&+L_3 \frac{4}{f^4_{\pi}} (-8m^4_K+s^2+t^2+4 m^2_K u) 
\nonumber \\&&-L_4 \frac{16 m^2_K 
u}{f^4_{\pi}}-L_5 \frac{8 m^2_K u}{f^4_{\pi}}+(2L_6+L_8) 
\frac{32 m^4_K}{f^4_{\pi}}
\end{eqnarray}

\noindent
The amplitude for $K^+ K^- \rightarrow K^0 {\bar K^0}$ is

\be
T_{cn,2}=\frac{2 m^2_K - u}{2 f^2_{\pi}}
\ee

\begin{eqnarray}
&T^T_{cn,4}&=\frac{A^r_{\pi}}{576 f^4_{\pi} {\pi}^2} (-4 m^2_K +m^2_{\pi} +6t)
+\frac{A^r_K}{288 f^4_{\pi} {\pi}^2} (10 m^2_K-3t) \nonumber \\ & &
+\frac{A^r_{\eta}}{576 f^4_{\pi} {\pi}^2} (20m^2_K-m^2_{\pi}-6u)
\end{eqnarray}

\begin{eqnarray}
&T^U_{cn,4}&=
\frac{24m^4_K-6m^2_{\pi} s-2s^2-2su+u^2+2m^2_K 
(4m^2_{\pi}-7t)}{384 f^4_{\pi} {\pi}^2}    \nonumber \\
&          &+\frac{A^r_{\pi}}{1152 f_{\pi}^4 \pi^2} 
(38m^2_K-2m^2_{\pi}-6s-15u) 
\nonumber \\&&+ \frac{A^r_K}{576 f^4_{\pi} {\pi}^2} (10 m^2_K+3s-6u)
+\frac{A^r_{\eta}}{1152 f^4_{\pi} {\pi}^2} (22m^2_K+
2m^2_{\pi}-15u)    \nonumber  \\
&          &+\frac{B^r_{\pi} (s)}{1536 f^4_{\pi} {\pi}^2} (8m^2_K 
(s-4m^2_{\pi})+s(7s-4t)+8m^2_{\pi} (s+2t))  \nonumber \\
&          &+\frac{B^r_{\pi}(t)}{384 f^4_{\pi} {\pi}^2} (s-u)(t-4m^2_{\pi})
+\frac{B^r_K (s)}{96 f^4_{\pi} {\pi}^2}(-8m^2_K+s(s-u)+
4m^2_K (s+u))  \nonumber \\
&          &+\frac{B^r_K (t)}{384 f^4_{\pi} {\pi}^2}(-8m^2_K+t(t-u)+
4m^2_K (t+u))+\frac{B^r_K (u)}{64 f^4_{\pi} {\pi}^2} (u-2m^2_K)^2 
\nonumber \\
&          &+\frac{B^r_{\eta} (s)}{4608 f^4_{\pi} {\pi}^2} 
(8m^2_K-9s)^2-\frac{B^r_{\pi\eta} (s)}{768 f^4_{\pi} {\pi}^2} (3s-4m^2_K)^2
\nonumber \\
&          &+\frac{B^r_{\pi\eta} (t)}{384 f^4_{\pi} {\pi}^2} (3t-4m^2_K)^2  
\end{eqnarray}

\begin{eqnarray}
&T^P_{cn,4}&=L_1 \frac{8}{f^4_{\pi}} (s-2m^2_K)^2 \nonumber \\ &&
+L_2 \frac{4}{f^4_{\pi}} (-8m^4_K+t^2+u^2+4m^2_K s) \nonumber \\
&          &+L_3 \frac{2}{f^4_{\pi}} (-8m^4_K+s^2+t^2+4m^2_K u)
\nonumber \\&&
+L_4 \frac{4}{3f^4_{\pi}} (-24 m^4_K+12 m^2_K s) -L_5 \frac{4 m^2_K u}
{f^4_{\pi}}\nonumber \\&&+(2L_6+L_8) \frac{16 m^4_K}{f^4_{\pi}}
\end{eqnarray}

\noindent
In the above formulas we have used the quantities

\be
A^r_P=-m^2_P \left[ -1 + \ln \left( \frac{m^2_P}{\mu^2} \right)
\right]
\label{defa}
\ee

\begin{eqnarray}
&B^r_{PQ}(s)&=\frac{\lambda^{1/2} (s,m^2_P,m^2_Q)}{2s} \, \ln \left(
\frac{m^2_P+m^2_Q-s+
\lambda^{1/2} (s,m^2_P,m^2_Q)}{m^2_P+m^2_Q-s-\lambda^{1/2} (s,m^2_P,m^2_Q)}
\right)
\nonumber  \\
&                &+2-\ln \left(\frac{m^2_Q}{\mu^2}\right)+
\frac{m^2_P-m^2_Q+s}{2s} \, \ln \left(\frac{m^2_Q}{m^2_P}\right)  
\label{defb}
\end{eqnarray}

\noindent
In the equal mass limit (\ref{defb}) reduces to 

\be
B^r_{P}(s)=2-\ln \, \left(\frac{m^2_P}{\mu^2}\right)-\sigma (m^2_P,s) \, \ln 
\, \left(\frac{\sigma (m^2_P,s)+1}{\sigma (m^2_P,s)-1}\right)
\ee

\noindent
where

\begin{eqnarray}
&\lambda(s,m^2_P,m^2_Q)&=[s-(m_P+m_Q)^2][s-(m_P-m_Q)^2]  \nonumber \\
&\sigma(m^2_P,s)       &=\sqrt{1-\frac{4m^2_P}{s}}
\label{sigma}
\end{eqnarray}

\noindent
The functions (\ref{defa}) and (\ref{defb}) come from the 
Passarino-Veltman integrals with one and two propagators \cite{PV}.

It is interesting to note that $L_7$ does not appear in $T_4^P$. 
This also happens in $\pi\pi \rightarrow \pi\pi$ and $K\pi\rightarrow 
K\pi$ ${\cal O} (p^4)$ scattering amplitudes.
$L_6$ and $L_8$ appear in the combination $2L_6 + L_8$ as a consequence of the 
Kaplan and Manohar symmetry \cite{K1}.

As it was stated before, the corrections coming from the ${\cal O} (p^4)$ 
calculation are, at least, as large as the lowest order contribution itself. 
For example in the $K {\bar K}$ threshold:

\begin{center}
$T_{cc,2}=56.5$
\end{center}

\begin{center}
$T_{cc,4}=73.6 + i \, 36.74$
\end{center}

\noindent
Unambiguously this means that a perturbative calculation is useless in this 
region and that some non perturbative scheme should be used in order to 
compare with the experimental phenomenology.

\vspace{0.5cm}
\noindent{\bf 3. \, Unitarization of the $\pi\pi$ and $K {\bar K}$ 
amplitudes}
\vspace{0.5cm}

We already stated in the introduction that we are going to use a unitarization 
procedure recently developed in \cite{O2} and thoroughly used and 
applied to phenomenology in 
\cite{O3}, where the S and P waves meson-meson amplitudes were reproduced 
successfully. However, in both works, the ${\cal O} (p^4)$ amplitudes were 
approximated. Here we are going to use our calculated ${\cal O} (p^4)$ 
$K {\bar K}$ 
amplitude together with the $\pi\pi$ and $K \pi$ ${\cal O} (p^4)$ scattering 
amplitudes given in \cite{M2}. 

In \cite{O2} a rather general scheme was derived to obtain final unitarized 
amplitudes from the ${\cal O} (p^2)$ and ${\cal O} (p^4)$ ChPT 
scattering amplitudes. 
For our case with two channels, labelled as 1 for $\pi\pi$ and 2 for 
$K \bar K$, the amplitude with definite $I,J$ is given by

\be
T^{(I,J)}=T^{(I,J)}_2 \cdot \left[ T^{(I,J)}_2-T^{(I,J)}_4 \right]^{-1}
 \cdot  T^{(I,J)}_2 
\label{unit0}
\ee

\noindent
where the different amplitudes are actually $2 \times 2$ matrices for
I=0, 1, according to the above labelling of the states, and just numbers for 
I=2.

Note that (\ref{unit0}) was obtained from an expansion of the $1/T$ matrix 
amplitude, which has a zero when $T$ has a pole. In this way one expects that 
the $1/T$ expansion in powers of masses and momenta is meaningful even above 
the resonance region, where the low energy expansion of $T$ has no sense

The projection in definite angular momentum is given by

\be
T^{(I,J)}=\frac{1}{32N\pi} \int^1_{-1} \, d(\cos \theta) \, T^{(I)} (s,t) \, 
P_J (\cos \theta)
\ee

\noindent
where $N=1$ for $K {\bar K} \rightarrow K {\bar K}$, $N=\sqrt{2}$ for $\pi\pi 
\rightarrow K {\bar K}$ and $N=2$ for $\pi\pi \rightarrow \pi\pi$, since in 
the isospin formalism the pions are identical particles.

With this normalization, in our case unitarity reads, for (I,J)=(0,0)

\begin{eqnarray}
&4m^2_{\pi}< s < 4m^2_{K}:& \qquad \hbox{Im} \, T_{11}= \sigma(m^2_{\pi},s) 
\left| T_{11} \right|^2    \\
&4m^2_K < s < 4m^2_{\eta}:& \qquad \hbox{Im} \, T_{12}=\sigma(m^2_{\pi},s) 
\, T_{11} T^*_{12}+\sigma(m^2_K,s) \, T_{12} T^*_{22}   \nonumber \\
&                         & \qquad
\hbox{Im} \, T_{22}= \sigma(m^2_{\pi},s) \left| T_{12} 
\right|^2+\sigma(m^2_K,s) \left| T_{22} \right|^2  \nonumber \\
&                         & \qquad
\hbox{Im} \, T_{11}=\sigma(m^2_{\pi},s) \left| T_{11} 
\right|^2+\sigma(m^2_K,s) \left| T_{12} \right|^2
\label{unit2}
\end{eqnarray}

\noindent
For (I,J)=(1,1)

\begin{eqnarray}
&4m^2_{\pi}< s < 4m^2_{K}:& \qquad \hbox{Im} \, T_{11}= \sigma(m^2_{\pi},s) \left| 
T_{11} \right|^2    \\
&4m^2_K < s :& \qquad \hbox{Im} \, T_{12}=\sigma(m^2_{\pi},s) 
\, T_{11} T^*_{12}+\sigma(m^2_K,s) \, T_{12} T^*_{22}   \nonumber \\
&                         & \qquad
\hbox{Im} \, T_{22}= \sigma(m^2_{\pi},s) \left| T_{12} 
\right|^2+\sigma(m^2_K,s) \left| T_{22} \right|^2  \nonumber \\
&                         & \qquad
\hbox{Im} \, T_{11}=\sigma(m^2_{\pi},s) \left| T_{11} 
\right|^2+\sigma(m^2_K,s) \left| T_{12} \right|^2
\label{unit3}
\end{eqnarray}

\noindent
And for (I,J)=(2,0)

\be
4m^2_{\pi}< s : \qquad \hbox{Im} \, T_{11}= \sigma(m^2_{\pi},s) \left| 
T_{11} \right|^2
\ee

\noindent
with $\sigma$ given by (\ref{sigma}).

\noindent
In matrix notation 
(\ref{unit2}) and (\ref{unit3}) become

\be
\hbox{Im} \, T=T \, \sigma \, T^*
\ee

\noindent
with $\sigma=\left( \sigma (m^2_{\pi},s), \sigma(m^2_K,s) \right)$
 a diagonal matrix.

In deriving (\ref{unit0}) in \cite{O2} one was concerned essentially with the right hand 
cut, responsible for unitarity in the corresponding channel. As a result, the 
imaginary part of the amplitudes to be used in (\ref{unit0}), above the 
lightest threshold (in our case the $\pi\pi$ one, $s=4m^2_{\pi}$) and 
below the highest one ($s=4m_K^2$), was 
restricted to come only 
from unitarity, (\ref{unit2}), neglecting the left hand cut contribution to 
the imaginary part that appears in $T_{cc,4}$ and $T_{cn,4}$ 
for $s < 4m^2_K-4m^2_{\pi}$.

However, we have maintained the left hand cut contribution to the imaginary 
part of $T_{cc,4}$ and $T_{cn,4}$ below the $K \bar K$ threshold.  
One way to see how large is the resulting deviation from unitarity is to 
check the value of the inelasticity in the energy region $4m^2_{\pi} < s < 
4m^2_K$ for (I,J)= (0,0) and (1,1) where two channels appear. In both cases 
the deviation from 1 is smaller than 1\%. 

The $\eta\eta$ intermediate state has been included 
in the ${\cal O} (p^4)$ ChPT 
amplitudes. However, for the (0,0) channel for $\sqrt{s} > 2m_{\eta}$ this 
state gives further contribution to the imaginary part of our amplitudes in 
addition to the one expressed in (\ref{unit2}). This means that (\ref{unit0}) 
with only the $\pi\pi$ and $K \overline K$ states does not fulfill 
unitarity strictly
for $\sqrt{s} > 2m_{\eta} \simeq$ 1.1 GeV. The influence of the $\eta\eta$
state is particularly
significative in the $K \overline K \rightarrow \pi\pi$ S-wave phase shifts, 
Fig. 6.
We will come back to this point later.

Omiting the $I,J$ labels, the relation between the $T$ and the 
$S$ matrix elements for a two channel process, in our case $\pi\pi$ 
and $K \overline K$ for (I,J)=(0,0) and (1,1), is given by

\begin{eqnarray}
&S_{11}=& 1+2i \, \sigma (m^2_{\pi},s) T_{11}    \nonumber \\
&S_{22}=& 1+2i \, \sigma (m^2_K,s) T_{22}   \nonumber \\
&S_{12}=& 2i \, \sqrt{\sigma(m^2_{\pi},s) \sigma(m^2_K,s)} T_{12}
\label{selem}
\end{eqnarray}

\noindent
To accomplish unitarity the $2 \times 2$ $S$-matrix can be written \cite{We1} 
as

\be
S=\left[\matrix{ \eta e^{2i\delta_1} & i(1-\eta^2)^{1/2} 
e^{i(\delta_1+\delta_2)} \cr  i(1-\eta^2)^{1/2} e^{i(\delta_1+\delta_2)} & 
\eta e^{2i\delta_2} } \right]
\label{smatrix}
\ee

\noindent
From (\ref{selem}) and (\ref{smatrix}) we obtain the phase shifts for $\pi\pi 
\rightarrow \pi\pi$ ($\delta_1$) and $\pi\pi \rightarrow K {\bar K}$ 
($\delta_1+\delta_2$) for $(I,J)=(0,0)$ and $(1,1)$.
For (I,J)=(2,0) only the $\pi\pi$ channel is necessary, when omiting multipion 
states. In this case, it is enough to consider 
the first equality of (\ref{selem}) to 
obtain the phase shifts with $S_{11}=e^{i2\delta}$.

\vspace{0.5cm}
\noindent{\bf 4. \, Fit, phase shifts and inelasticity} 
\vspace{0.5cm}

In ChPT the experimental values for the $L_i$ coefficients come from ${\cal 
O}(p^4)$ fits to low energy experimental data. Here we fit the $L_i$ 
constants to 
experiment in a much broader energy interval and with an expression valid 
to all orders. Hence, differences are expected between our fitted $L_i$ 
parameters and the values quoted from ChPT.  

Furthermore, our approach is not cross symmetric. This implies that 
contributions from the left hand cut of order higher than ${\cal O} (p^4)$ are 
effectively reabsorbed in the values of our $L_i$ coefficients. This point has 
been studied in \cite{Bo1} with the conclusion that the value 
of the $L_i$ obtained from a non cross symmetric method are influenced by 
this reabsorption procedure of the left hand cut. In this way, the value we 
quote for our $L_i$ constants has to be taken with care when comparing with 
the values of the $L_i$ from ChPT.

We have used simultaneously the phase shifts of the $\pi\pi \rightarrow 
\pi\pi$ with I=0 and 1, 
Figs. 4 and 5, to fit the value of our free parameters: 
$L_1,L_2,L_3,L_4,L_5$ and $2L_6+L_8$. The fit has been done using MINUIT.
In the energy region $\sqrt{s} =$ 500-950 MeV the data 
from different experiments for S-wave $\pi\pi$ phase shifts are incompatible.
Given that situation, we have taken as central value for each energy the mean 
value between the different experimental results [29-34]. For $\sqrt{s}=$ 
0.95-1 GeV, the mean value comes from \cite{Hya1,Gra1}. In both cases 
the error is the maximum between the experimental errors and the largest 
distance between the experimental points and the average value.

The quoted errors in the value we have obtained for the $L_i$ coefficients is 
just the statistical one.

The fit is pretty good, as can be seen in Figs. 4 and 5, with $\chi^2=1.3$ per 
degree of freedom.
The values we obtain at the $M_{\rho}$ scale and in units of $10^{-3}$ are 

\begin{eqnarray}
L_1&=&0.72^{+0.03}_{-0.02}     \nonumber  \\
L_2&=&1.36^{+0.02}_{-0.05}     \nonumber  \\
L_3&=&-3.24 \pm 0.04    \nonumber  \\
L_4&=&0.20 \pm 0.10     \nonumber  \\
L_5&=&0.0^{+0.8}_{-0.4}      \nonumber  \\
2L_6+L_8&=&0.00^{+0.26}_{-0.20}
\end{eqnarray}

\noindent
The small errors for $L_1,L_2$ and $L_3$ are due to the strong constraints 
imposed by 
the small errors in the experimental data of the $\delta_{11,\pi\pi}$ phase 
shift, Fig.~4.

The values of ChPT are

\begin{eqnarray}
L_1&=&  \, 0.4 \, \pm \, 0.3     \nonumber  \\
L_2&=&  \, 1.4 \, \pm \, 0.3     \nonumber  \\
L_3&=&  \, -3.5 \, \pm \, 1.1     \nonumber  \\
L_4&=&  \, -0.3 \, \pm \, 0.5     \nonumber  \\
L_5&=&  \, 1.4 \, \pm \, 0.5     \nonumber  \\
2L_6+L_8&=& \, 0.5 \, \pm \, 0.7    \nonumber  \\
\end{eqnarray}

\noindent

We can see that our values, taking errors into account, are compatible 
with those from ChPT, and then we 
can guarantee a good behaviour at low energies for our predictions.

Using these values for $L_i$ we also describe correctly phase shifts, 
scattering lengths and form factors for the $I=J=0$ and $I=J=1$ channels, 
as we will see later.

In Fig. 4 we show our fitted $\delta_1$ for $I=J=1$, from the two pion 
threshold up to 1.2 GeV and we see a good agreement with the experimental data 
which are dominated by the presence of the $\rho(770)$ that we reproduce 
nicely. In this channel we obtain that the influence of the $K {\bar K}$ 
channel coupled to $\pi\pi$ is negligible.

\begin{figure}[H]
\centerline{\protect\hbox{
\psfig{file=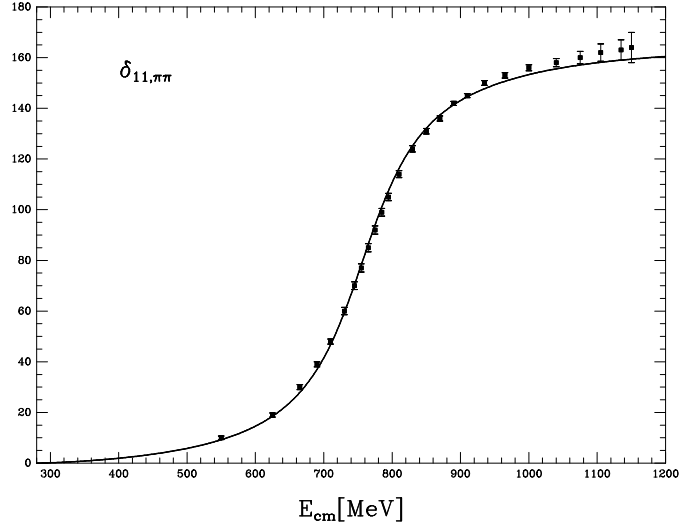,width=0.7\textwidth,angle=-90,silent=}}}
\caption{Phase shift for $\pi\pi \rightarrow \pi\pi$ in $I=J=1$. Data:
\cite{Pro1}.}
\label{11}
\end{figure}

\begin{figure}[H]
\centerline{\protect\hbox{
\psfig{file=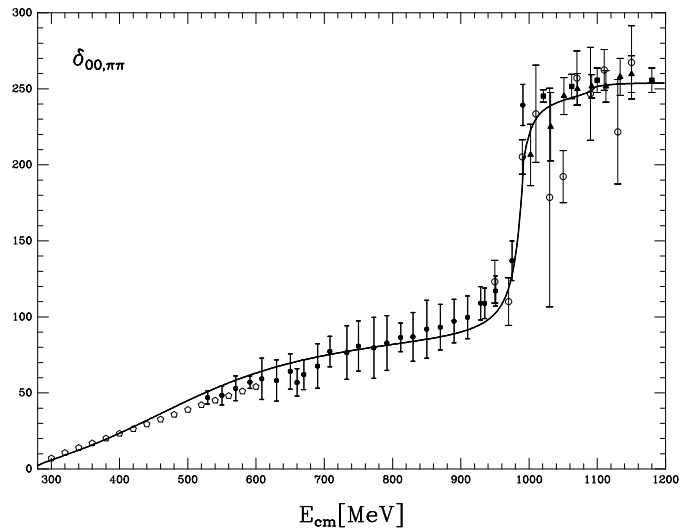,width=0.7\textwidth,angle=-90,silent=}}}
\caption{Phase shift for $\pi\pi \rightarrow \pi\pi$ in $I=J=0$. Data:
empty pentagon \cite{Fro1}, empty circle \cite{Kam1}, full square \cite{Gra1},
full triangle \cite{Hya1}, full circle represents the average explained
above.}
\label{00}
\end{figure}

In Fig. 5 we show $\delta_1$ for $I=J=0$, also from two pion threshold to 1.2 
GeV. The agreement with experiment is quite satisfactory showing 
clearly the presence of the $f_0 (980)$ resonance as a 
strong jump in the phase shift
 around 1 GeV. To get this resonance it is essential to include the kaons, and then to 
unitarize with coupled channel as we do. In this figure we also have plotted 
data from \cite{Fro1}, but since no error is quoted we have not included this 
data in the fit.

Now, once we have fixed the $L_i$ from the fit we predict other magnitudes.

\begin{figure}[H]
\centerline{\protect\hbox{
\psfig{file=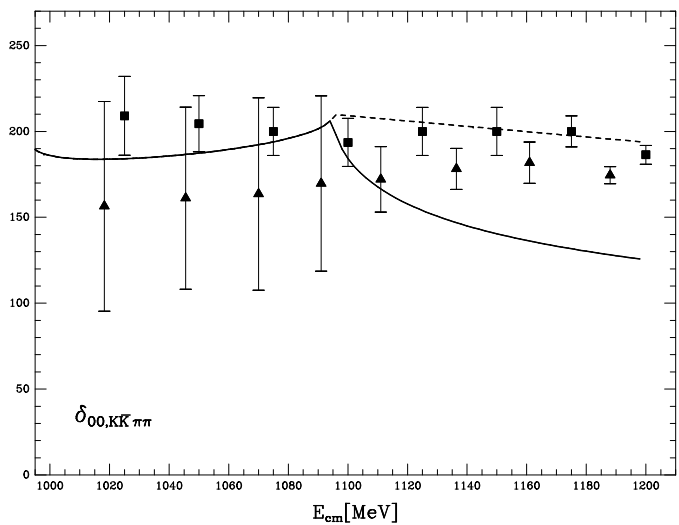,width=0.7\textwidth,angle=-90,silent=}}}
\caption{Phase shift for $\pi\pi \rightarrow K {\bar K}$ in $I=J=0$. Data:
full square \cite{Coh1}, full triangle \cite{Mar1}.}
\label{0012}
\end{figure}

In Fig. 6 the phase shift for the $K\bar{K}\rightarrow\pi\pi$ scattering, 
$\delta_1+\delta_2$, is shown for $I=J=0$. In this figure one sees clearly 
the $\eta\eta$ threshold. This process is the most sensible to the $\eta\eta$ 
intermediate state, contrary to what happens with the $\pi\pi$ phase shifts, Fig. 5, 
where its inclusion is almost negligible. One way to realize the influence of 
this channel in the former phase shifts is to cancel the imaginary part to $T_4$ 
coming from the intermediate $\eta\eta$ s-channel loop for $\sqrt{s}>2 m_{\eta}
\simeq 1.1$ GeV. In this way the dashed line curve in Fig. 6 is obtained, which agrees 
very well with data. This is telling us that the inclusion of the $\eta\eta$ channel 
in the unitarization procedure for the (0,0) channel is important for studying the 
$K\bar{K} \rightarrow \pi\pi$ scattering. In any case, as explained above, the threshold 
is very close to 1.2 GeV where other intermediate states, as four pions, are also 
important and should be included as well. Hence, we think that the inclusion of the 
$\eta\eta$ threshold in eq. (\ref{unit0}) should be done when going to higher energy, 
that is, when extending the model for energies higher than 1.2 GeV.

In Fig. 7, $(1-(\eta_{00})^2)/4$ is shown. Our results display the same 
tendency as the experimental data, particularly when taking into account the 
large experimental errors.

\vspace{0.8cm}

\begin{figure}[H]
\centerline{\protect\hbox{
\psfig{file=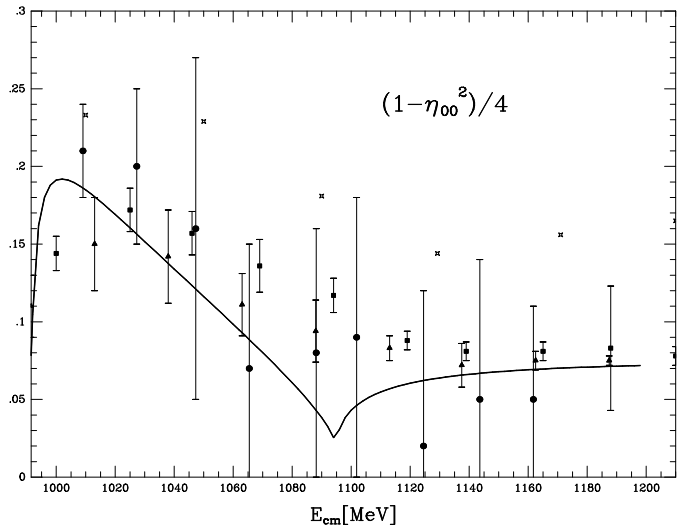,width=0.7\textwidth,angle=-90,silent=}}}
\caption{($1-(\eta_{00})^2)/4$, where $\eta_{00}$ is the inelasticity in
 $I=J=0$. Data: starred square \cite{Fro1}, full square \cite{Coh1}, full 
triangle \cite{Mar1}, full circle \cite{Och1}.}
\label{ine}
\end{figure}

\begin{figure}[ht]
\centerline{\protect\hbox{
\psfig{file=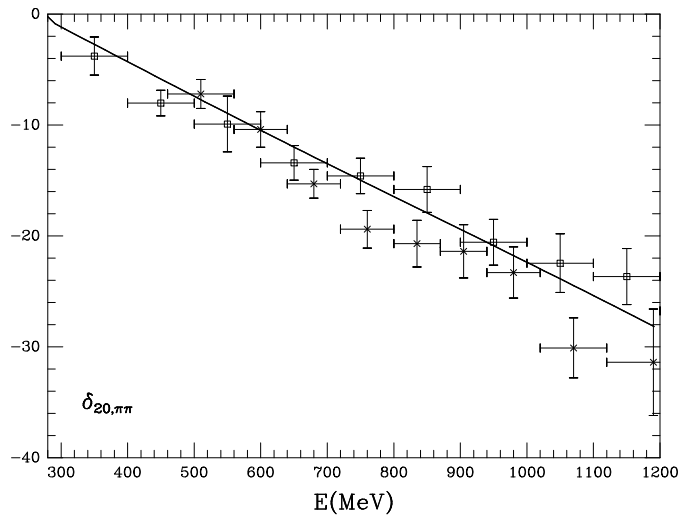,width=0.7\textwidth,angle=-90,silent=}}}
\caption{Phase shift for $\pi\pi \rightarrow \pi\pi$ in $I=2,J=0$. Data:
cross \cite{Ros1}, empty square \cite{Sch1}.}
\label{20}
\end{figure}

\vspace{0.8cm}

In Fig. 8 we show the $\pi\pi$ phase shift with I=2, J=0. The agreement with 
experimental data is fair. Contrary to the other channels we have shown, in 
this case no resonances appear and there is only the $\pi\pi$ channel.
So, in this case our result (apart of differences on the $L_i$ values) is the 
same than in the IAM \cite{D2}.

We have also calculated the scattering lengths for the three channels 
unitarized in this work, (I,J)=(0,0), (1,1) and (2,0). We denote them by 
$a^I_J$. In Table I we show the value we obtain for $a^I_J$  together with the 
experimental and the ChPT values to ${\cal O} (p^4)$. 
We see in this table that a good agreement 
with experiment is accomplished. Our values are also close to the ones from 
ChPT as one should expect because for low energies we recover the 
chiral~expansion.

\begin{center}
{\small{Table I: Comparison of scattering lengths in different channels}}
\end{center}

\begin{center}
\begin{tabular}{|c|c|c|c|}
\hline
\hline
$a^I_J$ & ChPT & Our results & Experiment\\
\hline
\hline
$a^0_0$ & $0.20 \pm 0.01$ & $0.210 \pm 0.002$ & $0.26 \pm 0.05$\\
$a^1_1$ & $0.037 \pm 0.002$ & $0.0356 \pm 0.0008$ & $0.038 \pm 0.002$\\
$a^2_0$ & $-0.041 \pm 0.004$ & $-0.040 \pm 0.001$ & $-0.028 \pm 0.012$\\
\hline
\end{tabular}
\end{center}

\vspace{0.5cm}
\noindent{\bf 5. \, Calculation of the scalar and vector pion form 
factors}
\vspace{0.5cm}

The scalar and vector form factors of the pion are defined respectively as

\be
\left<\pi^a (p') \pi^b (p) \, \hbox{out} \left| {\bar m} (\bar u u+ \bar d d) 
\right| 0  \right>=\delta^{ab} \Gamma(s)
\ee

\noindent
and
\be
\left< \pi^i (p') \pi^l (p) \, \hbox{out} \left| \bar q \gamma_{\mu} 
\left(\frac{\tau^k}{2}\right) q \right| 0 \right>=i \, \epsilon^{ikl} 
(p'-p)_{\mu} \, F_V (s)
\ee

\noindent
with $\bar m =(m_u+m_d)/2$ and $\epsilon^{ijk}$ the total antisymmetric tensor 
with three indices.

Assuming elastic unitarity (valid up to the $K \overline K$ threshold and 
neglecting multipion states) 
and making use of the Watson final state theorem 
\cite{Wa1} the phase of $\Gamma(s)$ and $F_V (s)$ is fixed to be the one of the 
corresponding partial wave strong amplitude:

\begin{eqnarray}
&\hbox{Im} \, \Gamma (s+i\epsilon)&= \tan \delta^0_0 \, \hbox{Re} \Gamma (s)  
\nonumber  \\
&\hbox{Im} \, F_V (s+i\epsilon)   &= \tan \delta^1_1 \, \hbox{Re} F_V (s)
\label{elasunit}
\end{eqnarray}
 
\noindent
The solution of (\ref{elasunit}) is well known and corresponds to the Omn\`es 
type \cite{Mu1,Om1}:

\begin{eqnarray}
&\Gamma (s)=& P_0 (s) \, \Omega_0 (s)   \nonumber \\
&F_V (s)   =& P_1 (s) \, \Omega_1 (s)
\label{ffactors}
\end{eqnarray}

\noindent
With
\be
\Omega_i (s)= \exp \left\{ \frac{s^n}{\pi} \int^{\infty}_{4m^2_{\pi}} \,
\frac{ds'}{{s'}^n} \, \frac{\delta^i_i (s')}{s'-s-i\epsilon} \right\}
\label{disp}
\ee

\noindent
In (\ref{ffactors}) $P_0 (s)$ and $P_1 (s)$ are polynomials of degree fixed by 
the number of 
subtractions done in $\ln \{ \Omega_0 (s) \}$ and $\ln \{ \Omega_1 (s)
\}$ minus one, and the zeros of $F_V$ and $\Gamma$.
For $n=1$, $P_i (s)=1$. This follows from the normalization requirement that 
$\Gamma (0)=F_V (0)=1$ and the absence of zeros for those quantities.

Since we have a prediction for the phase shifts we can calculate the 
dispersion integral (\ref{disp}) and obtain the pion form factors 
for both the scalar and 
vector cases. The results are shown in Figs. 9 and  10. The 
Omn\`es solution assumes the phase of the form factor to be that of the 
scattering amplitude, and that is true exactly only until the first inelastic 
threshold. 
The first inelastic threshold is the $4\pi$ one. However, as it was already 
said, its influence, in a first approach, is negligible. The first important 
inelastic threshold is the $K {\bar K}$ one around 1 GeV. This is essential in 
$I=J=0$ but negligible in $I=J=1$. This inelastic threshold, as 
discussed above, has been included in our approach and it is responsible for 
the appearance of the $f_0 (980)$ resonance, as it is clearly seen in Figs. 5 
and 10.

Due to the presence of this important $K {\bar K}$ threshold, overall for the 
$I=J=0$ sector, eqs. (\ref{ffactors}) are strictly correct up 
to $\sqrt{s}=2m_K$, 
which is indicated as a dashed-dotted vertical line in Figs. 9 and 10.
Up to this 
energy, we see the clear appearance in Figs. 9 and 10 of the $\rho(770)$ and 
$f_0 (980)$ respectively. In the case of the $F_V (s)$  the agreement with 
existing data is 
quite satisfactory. Above the $K {\bar K}$ threshold one expects deviations 
from (\ref{ffactors}) due to the opening of this inelastic channel. However, 
for the vector form factor we still see a rather good agreement with data and 
the deviation should be adscribed to the presence of the $\rho'$ resonance above 
1.2 GeV. On the other hand, the result obtained 
for the vector form factor is similar to the one recently obtained in 
\cite{F2} using another phase shift expression, taking into account possible 
uncertainties coming from orders higher than $p^4$ in ChPT. For the 
$I=J=0$ channel the most dramatic influence of the opening of the 
$K {\bar K}$ threshold is the appearance of the $f_0 (980)$  resonance, what
happens a little below $K {\bar K}$ threshold. Hence its appearance in 
Fig. 10 is well accommodated in our assumption of elastic unitarity for the 
Watson theorem, which we have used to evaluate the form factors.
So that we do not expect large deviations from our results 
even above the $K {\bar K}$ threshold up to the appearance of new $I=J=0$, 
$f_0$, resonances higher in energy, typically around $\sqrt{s} \sim$ 1.3 GeV.
In Fig. 10 the dashed line represents the 
scalar form factor unitarizing only with pions to obtain the $\delta_{00,
\pi\pi}$ phase shift, in the line of the works 
\cite{T1,T2,H1} and we see a very large 
influence of the $K {\bar K}$ channel through the $f_0$ resonance which is even 
substantial around $|\sqrt{s}|=500$ MeV.

\begin{figure}[ht]
\centerline{\protect\hbox{
\psfig{file=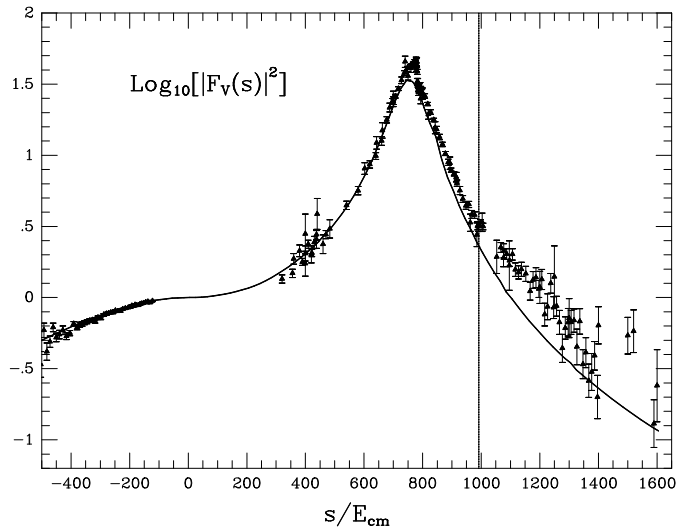,width=0.7\textwidth,angle=-90,silent=}}}
\caption{Vector pion form factor. The vertical line shows
the opening of the $K \overline K$ threshold. Data: \cite{Bar1}.}
\label{11f}
\end{figure}

\begin{figure}[H]
\centerline{\protect\hbox{
\psfig{file=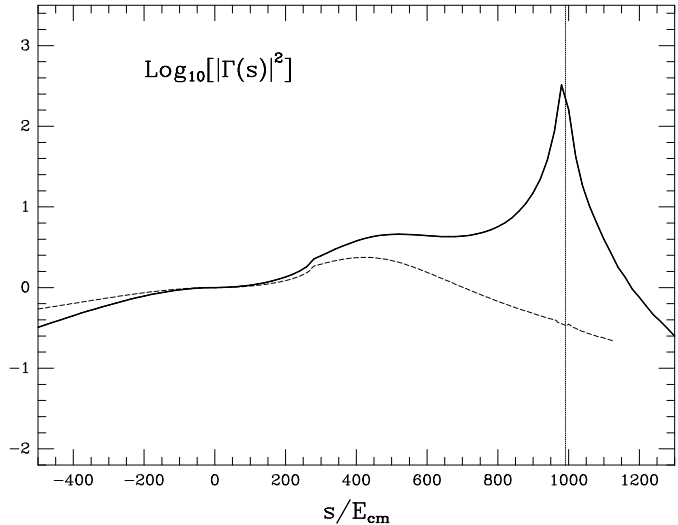,width=0.7\textwidth,angle=-90,silent=}}}
\caption{Scalar form factor. The dashed curve is the result unitarizing
only with pions. The solid line is the full result with both pions and kaons
in the intermediate state. The vertical line shows the opening
of the $K \overline K$ threshold.}
\label{00f}
\end{figure}

\vspace{0.5cm}
\noindent{\bf 6. Conclusions}
\vspace{0.5cm}

In this work we have calculated the $K {\bar K}$ scattering amplitude to next 
to 
leading order in ChPT. We have seen that from the $K \overline K$ threshold, 
due to the 
large kaon mass, very large corrections appear. This in principle should imply 
that this ${\cal O} (p^4)$ calculation is unlikely to be useful. 
However, we see 
that this is not true and that one can obtain accurate results when 
the ChPT calculations are supplied with some suitable non perturbative 
unitarization scheme 
as the one used here. In this way, we have successfully described the $\pi\pi 
\rightarrow \pi\pi$ phase shifts for (I,J)=(0,0), (1,1) and also the 
phase shift for 
$\pi\pi \rightarrow K {\bar K}$ in (I,J)=(0,0) and the inelasticity
in good agreement with experiment 
up to $\sqrt{s}=1.2$ GeV. 
The scattering lengths for $\pi\pi$ with I=0,1 and 2 are also calculated in 
agreement with experiment and former ChPT calculations.
We have finally computed the scalar and vector form 
factor of the pion making use of the above calculated $\pi\pi$ phase shifts 
in an Omn\`es representation and the results have also been satisfactory. 
The scalar form factor strongly shows that the $K {\bar K}$ channel is 
essential in order to reproduce the $f_0$ resonance.

\vspace{0.5cm}
\noindent
{\bf Acknowledgements}
\vspace{0.5cm}

We would like to acknowledge a critical reading and fruitful discussions 
with E. Oset and A. Pich. We also thank M.C. Gonz\'alez Garc\'{\i}a for 
her kind 
introduction to PAW and J. Fuster for his appreciated advices about the 
statistical behaviour of a fit.
The work of F. Guerrero and J. A. Oller has been supported 
by an FPI scholarship of the Spanish {\it Ministerio de Educaci\'on y Cultura} 
and of the {\it Generalitat Valenciana} respectively.

\end{document}